\documentclass[10pt,conference]{IEEEtran}
\usepackage{cite}
\usepackage{amsmath,amssymb,amsfonts}
\usepackage{algorithmic}
\usepackage{graphicx}
\usepackage{textcomp}
\usepackage{xcolor}
\usepackage{balance}
\usepackage{xspace}
\usepackage{pifont}
\usepackage{subcaption}
\usepackage{booktabs}
\usepackage{colortbl}
\usepackage{multirow}
\usepackage{url}

\def\BibTeX{{\rm B\kern-.05em{\sc i\kern-.025em b}\kern-.08em
    T\kern-.1667em\lower.7ex\hbox{E}\kern-.125emX}}

\definecolor{gray50}{gray}{.5}
\definecolor{gray40}{gray}{.6}
\definecolor{gray30}{gray}{.7}
\definecolor{gray20}{gray}{.8}
\definecolor{gray10}{gray}{.9}
\definecolor{gray05}{gray}{.95}

\newcommand{\ie}{\emph{i.e.,}\xspace}
\newcommand{\eg}{\emph{e.g.,}\xspace}
\newcommand{\etc}{etc.\xspace}

\begin{document}

\title{Characterizing and Detecting Mismatch in Machine-Learning-Enabled Systems}

\author{\IEEEauthorblockN{Grace A. Lewis, Stephany Bellomo, Ipek Ozkaya}
\IEEEauthorblockA{\textit{Carnegie Mellon Software Engineering Institute} \\
Pittsburgh, PA USA \\
\{glewis, sbellomo, ozkaya\}@sei.cmu.edu}
}

\maketitle

\begin{abstract}
Increasing availability of machine learning (ML) frameworks and tools, as well as their promise to improve solutions to data-driven decision problems, has resulted in popularity of using ML techniques in software systems.  However, end-to-end development of ML-enabled systems, as well as their seamless deployment and operations, remain a challenge. One reason is that development and deployment of ML-enabled systems involves three distinct workflows, perspectives, and roles, which include data science, software engineering, and operations. These three distinct perspectives, when misaligned due to incorrect assumptions, cause ML mismatches which can result in failed systems. We conducted an interview and survey study where we collected and validated common types of mismatches that occur in end-to-end development of ML-enabled systems. Our analysis shows that how each role prioritizes the importance of relevant mismatches varies, potentially contributing to these mismatched assumptions. In addition, the mismatch categories we identified can be specified as machine readable descriptors contributing to improved ML-enabled system development. In this paper, we report our findings and their implications for improving end-to-end ML-enabled system development. 
\end{abstract}

\begin{IEEEkeywords}
software engineering, machine learning, software engineering for machine learning, model engineering
\end{IEEEkeywords}

\section{Introduction}\label{sec:introduction}

Despite advances in frameworks for machine learning (ML) model development and deployment, integrating models into production systems still remains a challenge \cite{Bohm2019}\cite{Flaounas2017}\cite{Kriens2019}\cite{Lwakatare2019}\cite{Reisz2019}. One reason is that the development and operation of ML-enabled systems involves three perspectives, with three different and often completely separate workflows and people: the data scientist builds the model; the software engineer integrates the model into a larger system; and then operations staff deploy, operate, and monitor the system. These perspectives often operate separately, using different processes and vocabulary referring to similar concepts, leading to opportunities for mismatch between the assumptions made by each perspective with respect to the elements of the ML-enabled system, and the actual guarantees provided by each element. Examples of mismatch and their consequences include (1) poor system performance because computing resources required to execute the model are different from computing resources available during operations, (2) poor model accuracy because model training data is different from operational data, (3) development of large amounts of glue code because the trained model input/output is incompatible with operational data types, and (4) monitoring tools are not set up to detect diminishing model accuracy, which is the evaluation metric defined for the trained model.

We therefore define \textbf{ML Mismatch} as a problem that occurs in the development, deployment, and operation of an ML-enabled system due to incorrect assumptions made about system elements by different stakeholders (\ie data scientist, software engineer, operations) that results in a negative consequence. ML mismatch can be traced back to information that could have been shared between stakeholders that would have avoided the problem. 

The objective of our study is to develop a set of machine-readable descriptors for system elements, as a mechanism to enable mismatch detection and prevention in ML-enabled systems. The goal of the descriptors is to codify attributes of system elements in order to make all assumptions explicit. The descriptors can be used by system stakeholders to consistently document and share system attributes; and by automated mismatch detection tools at design time and run time for cases in which attributes lend themselves to automation. The research questions therefore defined for this study are \textbf{RQ1:} What are common types of mismatch that occur in the end-to-end development of ML-enabled systems?, \textbf{RQ2:} What are best practices for documenting data, models, and other system elements that will enable detection of ML mismatch?, and \textbf{RQ3:} What are examples of ML mismatch that could be detected in an automated way, based on the codification of best practices in machine-readable descriptors for ML system elements?

The focus of this paper is to report on the first phase of the study addressing RQ1, which is the results of practitioner interviews and a survey that were conducted to gather examples of real ML mismatches and their consequences. Section \ref{sec:study-design} presents the study design and Section \ref{sec:study-results} shows the results. In Section \ref{sec:analysis} we discuss findings and analysis insights. Section \ref{sec:towards-detection} outlines Phase 2 of our study on the path towards automated mismatch detection. Finally, Section \ref{sec:limitations} presents limitations, Section \ref{sec:related-work} talks about related work, and Section \ref{sec:conclusions} concludes the paper and presents next steps.

\section{Study Design}\label{sec:study-design}

We conducted 20 practitioner interviews to gather mismatch examples, and validated the interview results via a practitioner survey, as described in the following subsections.

\subsection{Interviews}\label{sec:interviews}

The goal of each interview was to gather examples of mismatch from practitioners in the roles of data scientist, software engineer, or operations for ML-enabled systems. Prior to each interview, each interviewee was sent a slide set describing the study. During the one-hour interview we followed a script to enable us to elicit examples of mismatch, consequences, and information that they believed that if shared would have avoided the mismatch. The presentation and interview guide are both available in the replication package for this study.\footnote{Replication package available at \url{github.com/GALewis/ML-Mismatch/}}

\begin{figure}[htbp]
  \centering
  \begin{subfigure}{0.49\columnwidth}
    \includegraphics[width=\textwidth]{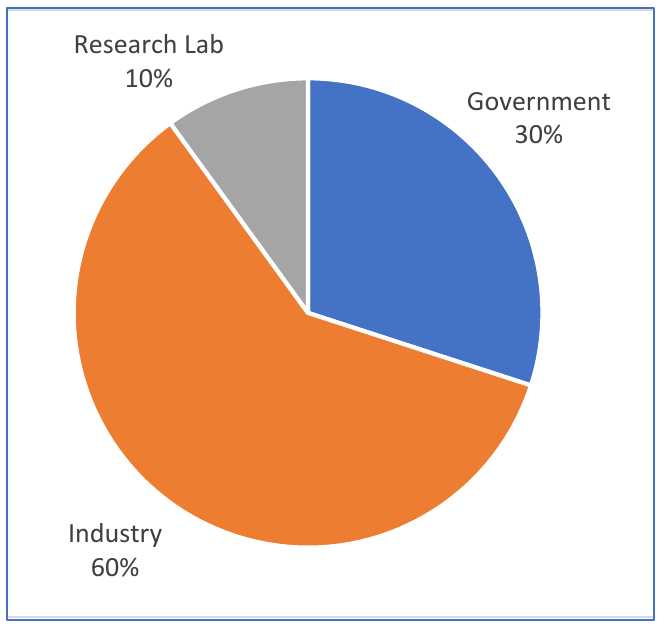}
    \caption{Affiliation} \label{fig:affiliation}
  \end{subfigure}%
  \hspace*{\fill}
  \begin{subfigure}{0.49\columnwidth}
    \includegraphics[width=\textwidth]{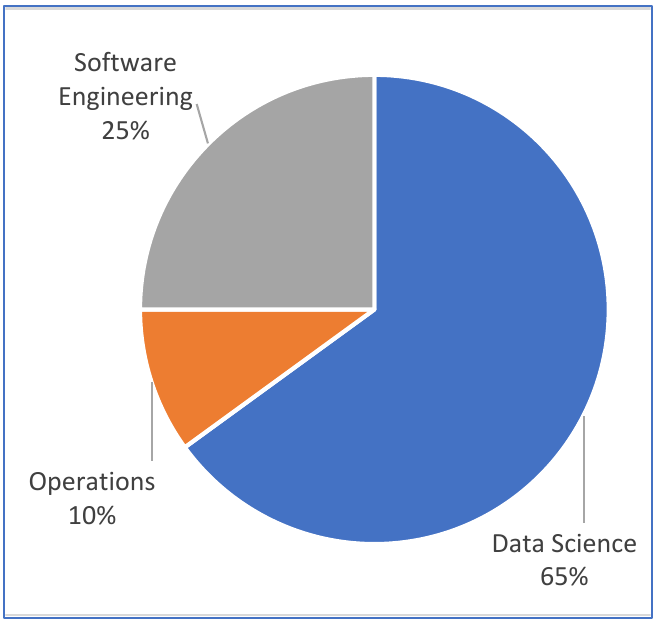}
    \caption{Primary Role} \label{fig:role}
  \end{subfigure}
  \hspace*{\fill}
  \caption{Interviewee Demographics} \label{fig:demographics}
\end{figure}

We conducted the interviews between November 2019 and July 2020. All the people interviewed were contacts from existing or previous collaborations or work engagements. We only interviewed people that confirmed to have experience developing or deploying operational ML-enabled systems. We did not interview people that only had academic model development experience or developed models that ran stand-alone to produce reports. Demographics for interviewees are presented in Figure \ref{fig:demographics}. 

Each interview was transcribed using a commercial transcription service and processed as follows. Steps were derived from well-established empirical software engineering guidelines \cite{Shull2007}\cite{Wohlin2012}.

\textit{Step 1: Transcript Segmentation:} Each transcript was imported into a spreadsheet by one researcher and divided into segments based on breaks in the interview that indicated when a new mismatch example was being discussed, \eg "What is another example of a mismatch that you experienced?"

\textit{Step 2: Mismatch Identification:} Each segment was evaluated independently by two researchers against the inclusion and exclusion criteria shown in Table \ref{tab:criteria} to determine if the segment exemplified a mismatch. A segment was identified as a mismatch example if it met all of the inclusion criteria and none of the exclusion criteria. We were only interested in mismatches resulting from information that was not explicitly stated and shared. For those identified as a mismatch, the researcher added a specific quote from the segment representative of the mismatch discussed, a short description, and the information that was not communicated that caused the mismatch.
    
\textit{Step 3: Mismatch Validation:} Results from the previous step were merged into a single spreadsheet. A third researcher reviewed each segment for which there was a disagreement and discussed with researchers until there was a consensus. For each agreed-upon mismatch, the third researcher produced a consolidated quote, short description, and information that should have been shared.
    
\textit{Step 4: Mismatch Coding:} After creating a spreadsheet with only the validated mismatches for all interviews, two researchers performed mismatch content analysis using open coding \cite{Lidwell2010} to categorize the types of mismatch identified in the interviews. The question that each researcher answered to perform the coding was “What element of an ML system does the information that was not communicated refer to?” The initial set of codes was created from the examples of ML system elements provided to interviewees as part of the project introduction slides.\footnote{Slide 4 in ML-Mismatch-Project-Introduction.pdf in replication package.} Codes were divided into major categories (\eg raw data, trained model, training data) and subcategories (\eg for trained model: programming language, API, version, \etc). The list of categories and sub-categories was expanded as researchers identified new codes. Given that mismatches could refer to more than one code, a researcher could assign up to three category/sub-category pairs per mismatch to ensure focus on the key mismatch causes. A third researcher consolidated codes after each round of open coding. Three rounds were conducted until agreement was reached.

\begin{table}[htbp] 
\caption{Inclusion and Exclusion Criteria}
\vspace{-2mm}
\centering
\scriptsize {
\begin{tabular}{ p{0.3cm} p{7.7cm} }
 \rowcolor{gray20}
 \multicolumn{2}{c}{\textbf{Inclusion Criteria}}\\\rowcolor{gray10} 
I1 & The segment describes a situation in which a system stakeholder made an assumption about a system element that was incorrect (\eg software engineer assumed that the model was ready to process operational data as-is). \\\rowcolor{gray05}
I2 & The situation described in the segment would not have occurred if information would have been shared between stakeholders (\eg data science team should have included the data pre-processing code along with the model). \\
\rowcolor{gray20}
\multicolumn{2}{c}{\textbf{Exclusion Criteria}}\\\rowcolor{gray10} 
E1 & The segment refers to problems that are internal to the data science process followed (\eg model parameters selected by the data scientist were not correct). \\ \rowcolor{gray05}
E2 & The segment refers to problems that are internal to the software engineering process (\eg different engineers were using different versions of Python). \\ \rowcolor{gray10} 
E3 & The segment refers to problems that are internal to the operations process (\eg the tool used for runtime monitoring did not have a good way to alert users of problems). \\ \rowcolor{gray05}
E4 & The situation described in the segment cannot be solved by sharing information between stakeholders (e.g., the data science team did not have enough data to train the model properly). \\ \rowcolor{gray10}
E5 & The segment refers to a statement that is not related to a mismatch example (\eg introductory statements, small talk, this is how we did version control in my previous job). \\
\end{tabular}
}
\label{tab:criteria}
\end{table} 

The resulting spreadsheet with anonymous results is included in the replication package. Note that interview transcripts are not included in the replication package per our protocol for human subject research (HSR) approved by our Institutional Review Board (IRB).

\subsection{Validation Survey}\label{sec:survey}

We conducted a survey to validate the resulting ML mismatch categories. In addition to demographics information, the survey asked each participant to rate the importance of sharing information related to each of the identified categories and subcategories for preventing mismatch. The participants were also given the opportunity to add any information that they considered was important but missing from the survey. The survey questions are also included in the replication package. To ensure that the survey reached participants who met the criteria of having experience developing or deploying operational ML-enabled systems, we sent the survey only to the interviewees and  asked them to share it with people in their organization according to the criteria. We used Qualtrics (\url{https://qualtrics.com}) as the survey administration tool. 

\section{Study Results}\label{sec:study-results}

\subsection{Interview Results}\label{sec:interview-results}

\begin{figure}[htbp]
  \centering
  \includegraphics[width=\linewidth]{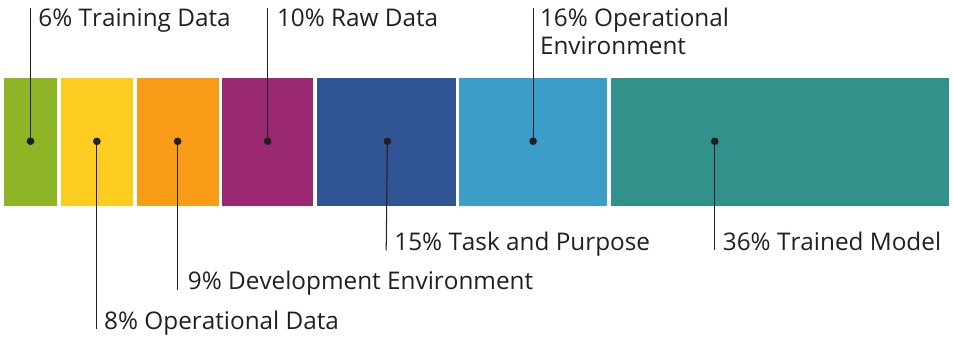}
  \caption{Mismatch Categories} 
  \label{fig:mismatch-categories}
\end{figure}

\begin{figure*}[htbp]
  \centering
  \includegraphics[width=15cm]{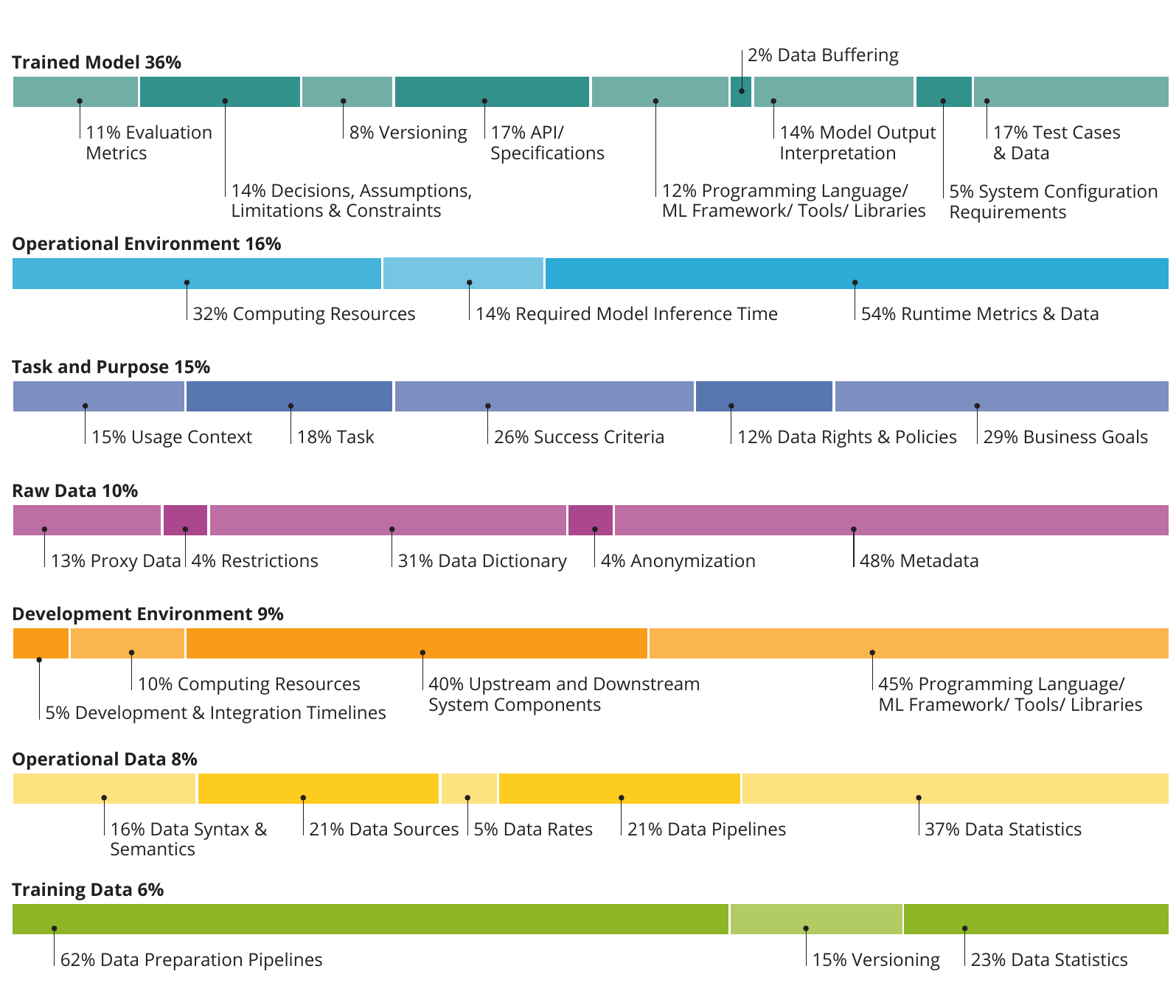}
  \caption{Mismatch Subcategories} 
  \label{fig:mismatch-subcategories}
\end{figure*}

A total of 140 mismatches were identified in the interviews, which resulted in 232 instances of information that was not communicated that led to mismatch. The resulting mismatch categories based on open coding and their occurrence are presented in Figure \ref{fig:mismatch-categories}. Each category is divided into subcategories, which are shown in Figure \ref{fig:mismatch-subcategories}, along with their occurrence in each category. 

Most identified mismatches refer to incorrect assumptions about the \textit{Trained Model} (36\%), which is the model trained by data scientists that is passed to software engineers for integration into a larger system. The next category is \textit{Operational Environment} (16\%), which refers to the computing environment in which the trained model executes (\ie model-serving or production environment). Categories that follow are \textit{Task and Purpose} (15\%) which are the expectations and constraints for the model, and \textit{Raw Data} (10\%) which is the operational or acquired data from which training data is derived. Finally, in a smaller proportion are the \textit{Development Environment} (9\%) used by software engineers for model integration and testing, the \textit{Operational Data} (8\%) which is the data processed by the model during operations, and the \textit{Training Data} (6\%) used to train the model.

\textbf{Trained Model (TM).} Most mismatches were related to lack of test cases and data that could be used for integration testing (17\%); and lack of model specifications and APIs (17\%). One software engineer interviewee stated \textit{"I was never able to get from the [data scientists] a description of what components exist, what are their specifications, what would be some reasonable test we could run against them so we could reproduce all their results.”} Other subcategories included unawareness of decisions, assumptions, limitations, and constraints that affect model integration and deployment (14\%); information necessary to interpret model output, results, or inferences (14\%); programming language, ML framework, tools, and libraries used in the development and training of the model (12\%);  evaluation metrics and results of trained model evaluation such as false positive rate, false negative rate, and accuracy (11\%); version information (8\%); system configuration requirements for trained model to execute, such as number of CPUs and GPUs, libraries, tools, and dependencies (5\%); and data buffering or time window requirements that would indicate that data has to be delivered in "chunks" instead of streamed (2\%).

\textbf{Operational Environment (OE).} Most mismatches were associated with lack of runtime metrics, logs, user feedback, and other data collected in the operational environment to help with troubleshooting, debugging, or retraining (54\%). One data scientist interviewee stated \textit{"A typical thing that might happen is that in the production environment, something would happen.  We would have a bad prediction, some sort of anomalous event. And we were asked to investigate that.  Well, unless we have the same input data in our development environment, we can't reproduce that event."} Other subcategories were unawareness of computing resources available in the operational environment, such as CPUs, GPUs, memory, and storage (32\%); and required model inference time (\ie time for the model to produce a result) (14\%).

\textbf{Task and Purpose (TP).} Most mismatches were related to lack of knowledge of business goals or objectives that the model was meant to satisfy (29\%). One data scientist interviewee stated \textit{"It feels like the most broken part of the process because the task that comes to a data scientist frequently is – hey, we have a lot of data. Go do some data science to it ... And then, that leaves a lot of the problem specification task in the hands of the data scientist."} Other subcategories included unawareness of success criteria, client expectations, validation scenarios, or acceptance criteria for determining that the model is performing correctly (26\%); task that model is expected to perform (18\%); how the results of the model are going to be used by end users or in the context of a larger system (15\%); and known data rights, legal, privacy, and other policies that need to be met by model and data (12\%).

\textbf{Raw Data (RD).} Most mismatches were associated with lack of metadata about raw data, such as how it was collected, when it was collected, distribution, geographic location, and time frames (48\%); and lack of a data dictionary that describes field names, description, values, and meaning of missing or null values (31\%). One data scientist interviewee stated \textit{"Whenever they had data documentation available, that was amazing because you can immediately reference everything, bring it together, know what's missing, know how it all relates. In the absence of that, then it gets incredibly difficult because you never know exactly what you're seeing, like is this normal?  Is it not normal?  Can I remove outliers? What am I lacking?  What do some of these categorical variables actually mean?”} Other subcategories were unawareness of the process used to generate or acquire proxy data due to sensitivities, legal, or policy reasons (13\%); indication regarding data sensitivities that would prohibit for example upload to public cloud environments (4\%); and the process used to anonymize data due to personally identifiable information constraints (4\%).

\textbf{Development Environment (DE).} Most mismatches were related to lack of knowledge of programming languages, ML frameworks, tools, and libraries in the development environment (45\%). One software engineer interviewee stated \textit{"The weird failures that you see porting models from R prototypes to other languages is interesting … almost like re-optimizing the whole model for a new language … I was able to diagnose that the way floating point numbers are handled in R and Python does not translate directly.”} Other subcategories include unawareness of specifications or APIs for upstream and downstream components (40\%); computing resources available in the development environment (10\%); and development and integration timelines for the larger system (5\%).

\textbf{Operational Data (OD).} Most mismatches were associated with lack of operational data statistics that could be used by data scientists to validate appropriateness of training data (37\%); and details on the implementation of data pipelines for the deployed model (21\%). One operations interviewee stated \textit{"There's the data inputs being restructured appropriately on the prototypes with this big complicated data pipeline leading up to them ... and we take it to deployment and you don't have the data coming through that same route anymore.  You want to have it being straight from the sensor data. If they reconstruct that pipeline onboard ... there's so many opportunities there for mismatches."} Other subcategories were unawareness of sources for operational data (21\%); syntax and semantics of the data that constitutes the input for the operational model (16\%); and rates at which operational data feeds into the operational model (5\%).

\textbf{Training Data (TD).} Most mismatches were related to lack of details of data preparation pipelines to derive training data from raw data (62\%). One software engineer interviewee stated \textit{"A group developed the architecture for a whole ML pipeline … but as a consequence of that, I think they sort of went a few steps further than they should have, creating lock-in, and kind of took over the feature engineering phase as well ... The mismatch was really at the design phase of the architecture of the machine learning pipeline where it really precluded us from doing more extensive research into alternative model architectures."} Other subcategories include unawareness of training data statistics (23\%); and version information (15\%).

\subsection{Validation Survey Results}\label{sec:survey-results}

\begin{figure}[htbp]
  \centering
  \begin{subfigure}{0.49\columnwidth}
    \includegraphics[width=\textwidth]{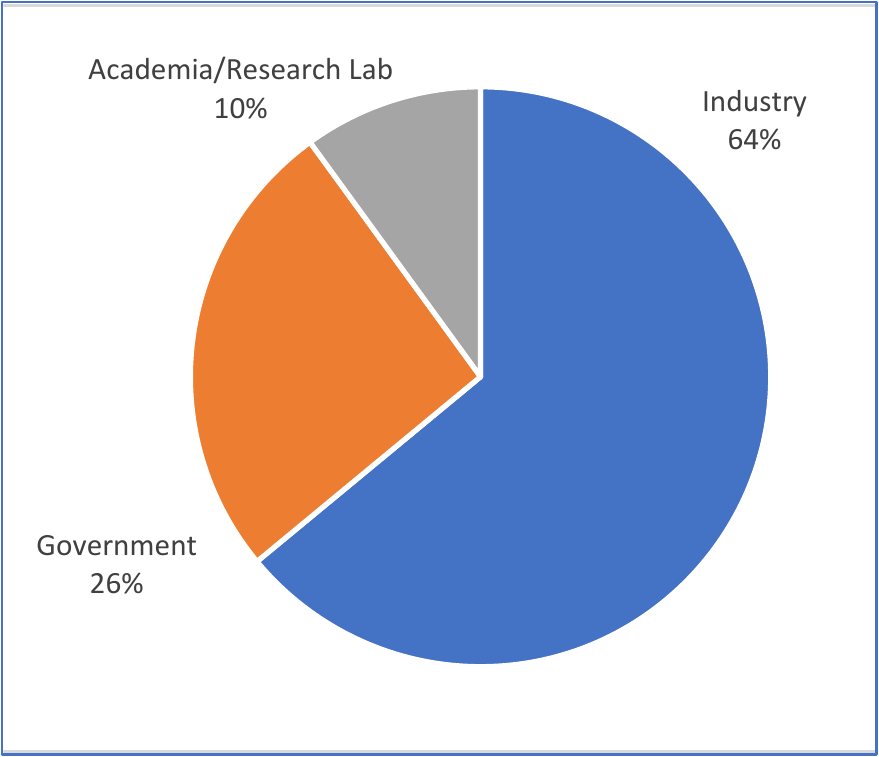}
    \caption{Affiliation} \label{fig:affiliation-survey}
  \end{subfigure}%
  \hspace*{\fill}
  \begin{subfigure}{0.49\columnwidth}
    \includegraphics[width=\textwidth]{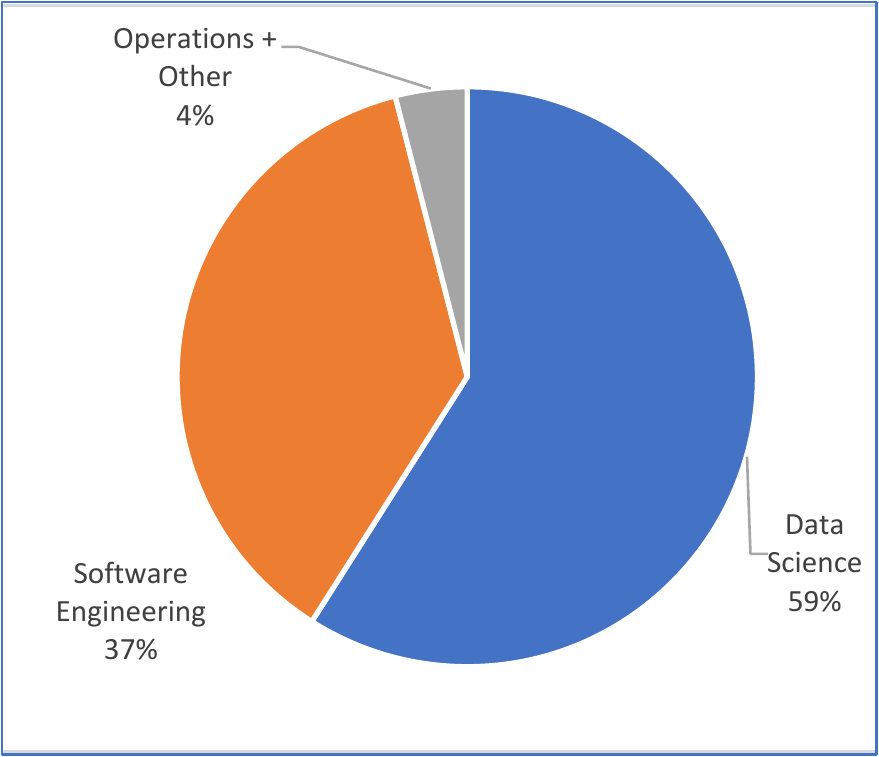}
    \caption{Primary Role} \label{fig:role-survey}
  \end{subfigure}
  \hspace*{\fill}
  \begin{subfigure}{0.49\columnwidth}
    \includegraphics[width=\textwidth]{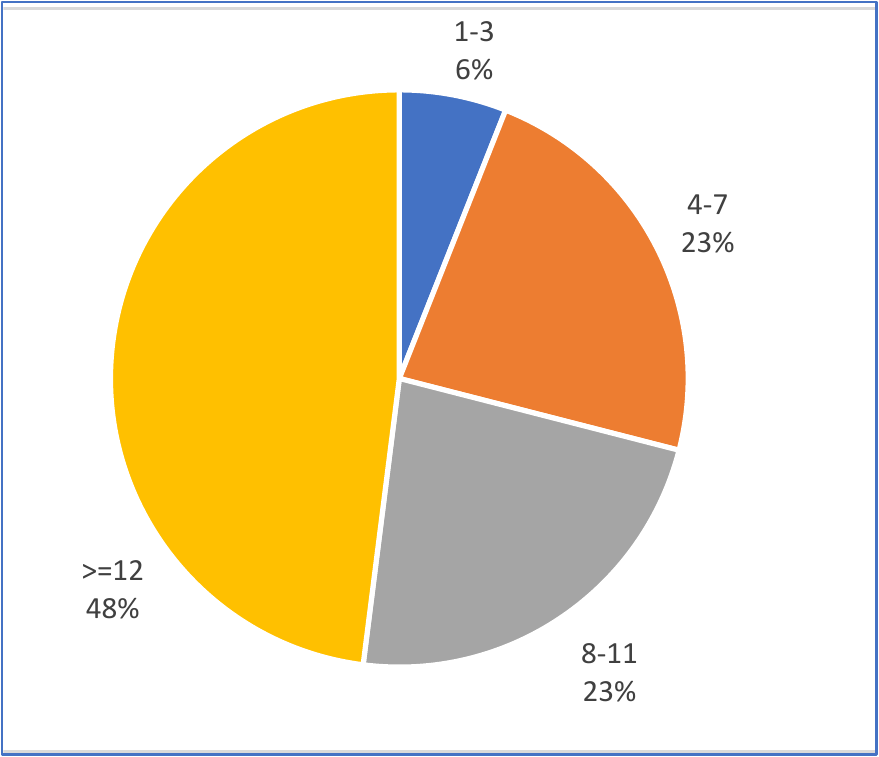}
    \caption{Total Years of Experience} \label{fig:experience-survey}
  \end{subfigure}%
  \hspace*{\fill}
  \begin{subfigure}{0.49\columnwidth}
    \includegraphics[width=\textwidth]{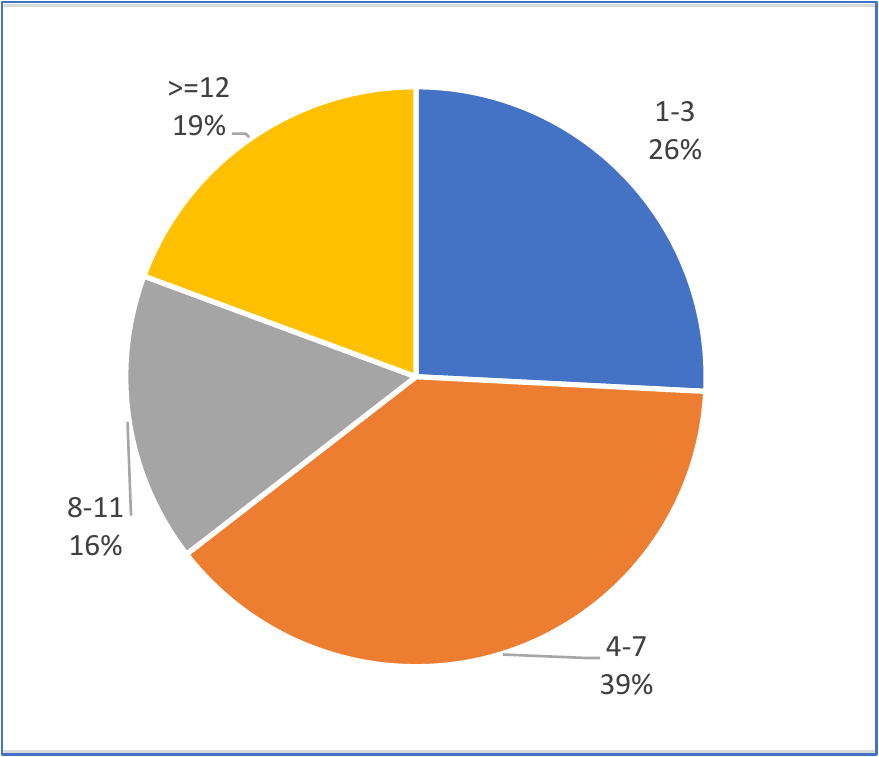}
    \caption{ML Years of Experience} \label{fig:ml-experience-survey}
  \end{subfigure}
  \hspace*{\fill}  
  \caption{Survey Respondent Demographics} \label{fig:survey-demographics}
\end{figure}

\begin{figure}[htbp]
  \centering
  \includegraphics[width=\linewidth]{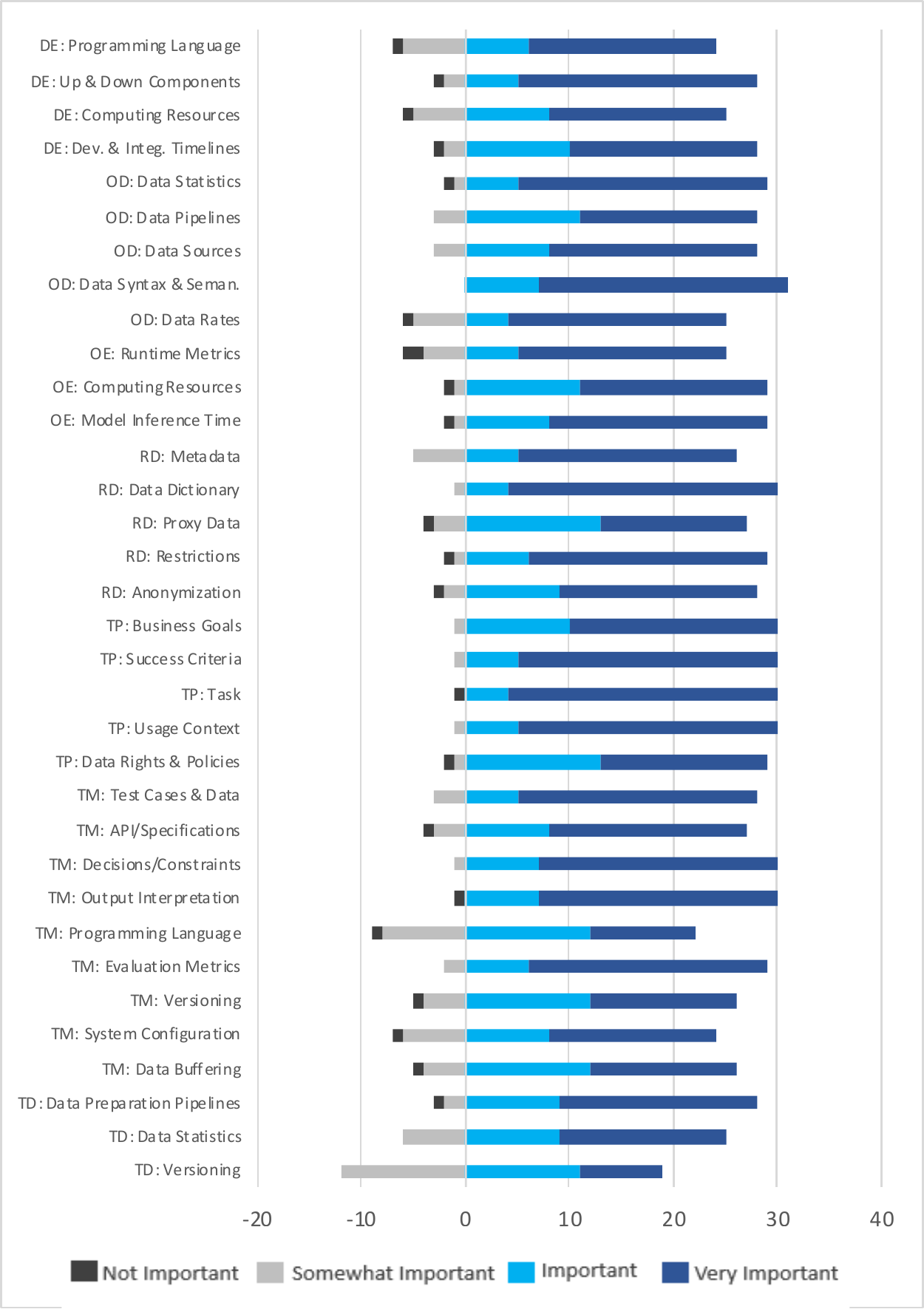}
  \caption{Survey Responses} 
  \label{fig:survey-all}
\end{figure}

A total of 31 survey responses were collected (15 from interviews), and are included in the study replication package. Survey demographics are shown in Figure \ref{fig:survey-demographics}. We recognize the small number of respondents in the \textit{Operations} role as a limitation, which is why our analyses will focus mostly on the \textit{Data Science} and \textit{Software Engineering} roles. However, we also highlight that because we were  specifically targeting practitioners, we asked our original interviewees to help us identify people in all three roles, and in most cases they could not identify an \textit{Operations} person on their team. While simply a conjecture, the fact that \textit{Operations} staff are not considered a key stakeholder in the end-to-end development and evolution of ML-enabled systems indicates the general lack of understanding of the key role of operations, and especially runtime monitoring, in these types of systems. For reporting purposes we combined \textit{Operations} with the \textit{Other} category, which were respondents who are currently in management-related roles. Results for all responses are shown in Figure \ref{fig:survey-all}. As shown in this figure, the importance of sharing information related to each subcategory to avoid mismatch was mostly rated between \textit{Important} and \textit{Very Important} for all, which demonstrates the validity of the identified causes for mismatch. 
 
\section{Analysis and Discussion}\label{sec:analysis}

In this section we analyze some of the interview and survey results, and discuss implications for software engineering practices and tools when developing ML-enabled systems. While there are many observations that we could make about the data, we limit our analysis to those that inform software engineering best practices and tools. Numbers reported are extracted from Figures \ref{fig:mismatch-subcategories} and \ref{fig:survey-all} and Table \ref{tab:data-per-role}. 

\begin{table}
    \vspace{-2mm}
    \centering
    \caption{Mismatch Categories Rated Very Important or Important (VI+I) per Role}
    \scriptsize {
    \begin{tabular}{|l|c|c|c|c|c|c|}
    \hline
         & \multicolumn{2}{c|}{\textbf{Data}} & \multicolumn{2}{c|}{\textbf{Software}}  & \multicolumn{2}{c|}{\textbf{Operations}} \\
         & \multicolumn{2}{c|}{\textbf{Science}} & \multicolumn{2}{c|}{\textbf{Engineering}}  & \multicolumn{2}{c|}{\textbf{+ Other}} \\ \hline
         & \textbf{VI+I} & \textbf{\%} & \textbf{VI+I} & \textbf{\%} & \textbf{VI+I} & \textbf{\%} \\ \hline
        DE: Programming Language & 11 & 69 & 8 & 80 & 5 & 100 \\ \hline
        DE: Up \& Down Components & 14 & 88 & 10 & 100 & 4 & 80 \\ \hline
        DE: Computing Resources & 11 & 69 & 9 & 90 & 5 & 100 \\ \hline
        DE: Dev. \& Integ. Timelines & 14 & 88 & 9 & 90 & 5 & 100 \\ \hline
        OD: Data Statistics & 15 & 94 & 10 & 100 & 4 & 80 \\ \hline
        OD: Data Pipelines & 15 & 94 & 8 & 80 & 5 & 100 \\ \hline
        OD: Data Sources & 14 & 88 & 9 & 90 & 5 & 100 \\ \hline
        OD: Data Syntax \& Seman. & 16 & 100 & 10 & 100 & 5 & 100 \\ \hline
        OD: Data Rates & 11 & 69 & 10 & 100 & 4 & 80 \\ \hline
        OE: Runtime Metrics & 13 & 81 & 9 & 90 & 3 & 60 \\ \hline
        OE: Computing Resources & 14 & 88 & 10 & 100 & 5 & 100 \\ \hline
        OE: Model Inference Time & 14 & 88 & 10 & 100 & 5 & 100 \\ \hline
        RD: Metadata & 14 & 88 & 9 & 90 & 3 & 60 \\ \hline
        RD: Data Dictionary & 16 & 100 & 10 & 100 & 4 & 80 \\ \hline
        RD: Proxy Data & 14 & 88 & 9 & 90 & 4 & 80 \\ \hline
        RD: Restrictions & 15 & 94 & 10 & 100 & 4 & 80 \\ \hline
        RD: Anonymization & 14 & 88 & 9 & 90 & 5 & 100 \\ \hline
        TP: Business Goals & 16 & 100 & 9 & 90 & 5 & 100 \\ \hline
        TP: Success Criteria & 15 & 94 & 10 & 100 & 5  & 100 \\ \hline
        TP: Task & 15 & 94 & 10 & 100 & 5 & 100 \\ \hline
        TP: Usage Context & 15 & 94 & 10 & 100 & 5 & 100 \\ \hline
        TP: Data Rights \& Policies & 15 & 94 & 10 & 100 & 4 & 80 \\ \hline
        TM: Test Cases \& Data & 14 & 88 & 10 & 100 & 4 & 80 \\ \hline
        TM: API/Specifications & 14 & 88 & 9 & 90 & 4 & 80 \\ \hline
        TM: Decisions/Constraints & 15 & 94 & 10 & 100 & 5 & 100 \\ \hline
        TM: Output Interpretation & 15 & 94 & 10 & 100 & 5 & 100 \\ \hline
        TM: Programming Language & 11 & 69 & 7 & 70 & 4 & 80 \\ \hline
        TM: Evaluation Metrics & 16 & 100 & 8 & 80 & 5 & 100 \\ \hline
        TM: Versioning & 14 & 88 & 8 & 80 & 4 & 80 \\ \hline
        TM: System Configuration & 12 & 75 & 7 & 70 & 5 & 100 \\ \hline
        TM: Data Buffering & 14 & 88 & 8 & 80 & 4 & 80 \\ \hline
        TD: Data Preparation Pipelines & 15 & 94 & 9 & 90 & 4 & 80 \\ \hline
        TD: Data Statistics & 13 & 81 & 9 & 90 & 3 & 60 \\ \hline
        TD: Versioning & 12 & 75 & 5 & 50 & 2 & 40 \\ \hline
    \end{tabular}
    }
    \label{tab:data-per-role}
\end{table}

Most mismatches identified during the interviews are related to incorrect assumptions about the \textit{Trained Model} (36\%). This is not surprising because the model constitutes the main "hand off" from a data science team to a software engineering team. Within this category, most mismatches were related to \textit{Test Cases and Data} and \textit{API/Specifications}, which are two pieces of information that are key for model integration into a larger system. However, survey data shows that what is most important to share about the trained model varies for each role. For the Data Scientist it is \textit{Evaluation Metrics} because they (1) set expectations for model performance and (2) establish a baseline for runtime monitoring of model performance over time. What is most important for the Software Engineer is a tie between \textit{Test Cases and Data}, \textit{Decisions/Assumptions/Limitations/Constraints}, and \textit{Model Output Interpretation}. For a software engineer these two last pieces of information provide insights into (1) any additional components that needs to be developed to address incompatibilities and (2) how to properly pass model results to downstream components. For \textit{Operations + Others}, \textit{System Configuration} is one of the most important categories, but it is not for the other roles. This is an expected result because Operations is responsible for serving the model and meeting any established service-level agreements (SLAs). It also hints at the fact that the operational environment might not usually be considered a constraint for model development, which leads to mismatches identified in the interviews in which complex models are created that cannot be served in the operational environment.

Specifically related to \textit{Test Cases \& Data} there were 14 mismatches. In some cases, software engineer interviewees reported receiving test data and results from data scientists that could be used for model integration testing. However, they also reported that test data used for model development was often also the cause for mismatch because it is not enough for generating appropriate test cases; specificially it (1) does not take into account the often uncertain nature of ML models, \ie output should be expressed as acceptable boundaries or an expected order of results instead of exact values, and (2) does not include error cases such as input errors and out-of-distribution (OOD) data. While the concept of test cases is common for software engineering, it is not common for data scientists, which shows the value of having a shared understanding between different perspectives and roles.

\textit{Data Syntax \& Semantics} for \textit{Operational Data} rated most important in surveys. However, there were only three mismatch examples related to this subcategory in our interview data. This could be because for many systems the \textit{Raw Data} comes from the \textit{Operational Data} and therefore is already well-known or documented, in which case no mismatch was observed.  Further collection of mismatch examples would be needed to better understand this type of mismatch in practice. 

With respect to the \textit{Operational Environment}, \textit{Runtime Metrics} received the most \textit{Not Important} responses in the survey, yet in the interviews runtime metrics had the largest number of mismatch examples within this category (20). Runtime metrics in ML-enabled systems are critical for continuous improvement of the model, especially to detect model drift and any other indicators that it is time to retrain and redeploy the model. We attribute this disconnect to the fact that the survey had the lowest number of respondents from the Operations role. ML-enabled system development still does not have agreed-upon, end-to-end development practices and most of the attention is currently in model development and not informed evolution, which can also explain the variance. To quote one of our survey respondents \textit{"Mismatch in understanding what it means to run successful under real-world conditions is the \#1 mismatch ... Some of the biggest mismatches in operational environment have to do with how a production system handles failure and overload."} We envision runtime metrics comprising \textit{algorithm metrics} related to data drift (\eg differences in data distribution often referred to as training-serving skew, non-expected inputs) and \textit{model performance metrics} (\eg accuracy, false positive and false negative rates, user feedback). This would require agreement between trained model evaluation metrics and operational environment runtime metrics to ensure feasibility and completeness. Agreement on what information to log is also important (\eg should logging be done for all input/output pairs or only for anomalies).

Most mismatch subcategories under \textit{Task and Purpose} were rated as \textit{Very Important} across all roles.  We gathered 34 examples of mismatch caused by not having a shared understanding of what basically constitutes the requirements for the model. As stated by one of the mismatch examples collected in our survey: \textit{"It is key to understand the problem being solved ... It is easy to get trapped tuning a model that doesn't actually solve the problem."}  We envision model requirements to include business requirements comprising the subcategories listed under \textit{Task and Purpose}, in addition to technical requirements such as the subcategories listed under \textit{Development Environment} and \textit{Operational Environment} in Figure \ref{fig:mismatch-subcategories}.

In general, the mismatch subcategory that was considered least important by both interviewees and all survey respondent roles was \textit{Training Data: Version}. This was a surprising result because of the tight relationship between model performance and training data. For model troubleshooting and retraining, knowing the exact data that was used to train the deployed model would seem important. The other subcategories in \textit{Training Data} were also generally rated low in importance even though the interviews showed several negative consequences of not having this information, such as the inability to perform data drift detection and other runtime monitoring when \textit{Distribution} and \textit{Data Statistics} are not known.  In addition, when details of \textit{Data Preparation Pipelines} are not known there is lack of clarity of how much data manipulation and validation happens inside the model code and how much happens outside; this subcategory was rated of higher importance by software engineers, likely because they have to deal with the consequences of not having this information.  

\textit{Metadata} in the \textit{Raw Data} category related to 11 mismatch examples and was considered \textit{Important/Very Important} by 81\% of survey respondents. \textit{Data Dictionary} was related to 7 mismatch examples and was considered \textit{Important/Very Important} by 94\% of survey respondents. Using data dictionaries is a practice that is common in the database community that not surprisingly would be well received in the data science community to better understand raw data. However, having access to metadata provides insights into how representative data is of operational data, which is equally or even more important, which our survey results do not reflect as strongly.

\textit{Programming Language/ML Framework/Tools/Libraries} was a subcategory that contained a large number of interview mismatch examples in both \textit{Development Environment} (9) and \textit{Trained Model} (10), which can be seen as counterparts. Most of these mismatches had to do with having to port models because of language differences, which when combined with lack of model API/Specifications is a very error-prone activity. However, these two categories were not as important among survey respondents. One explanation is that perhaps these respondents did not have to deal with model porting, which seemed to be common in interviewees mostly from Government, in which models are developed by contractors outside of their organizations. Another explanation is a growing trend towards deploying models as microservices, which would in fact hide some of these differences \cite{Sato2019}.

\section{Towards Automated Mismatch Detection}\label{sec:towards-detection}

\begin{figure}[htbp]
  \centering
  \includegraphics[width=\linewidth]{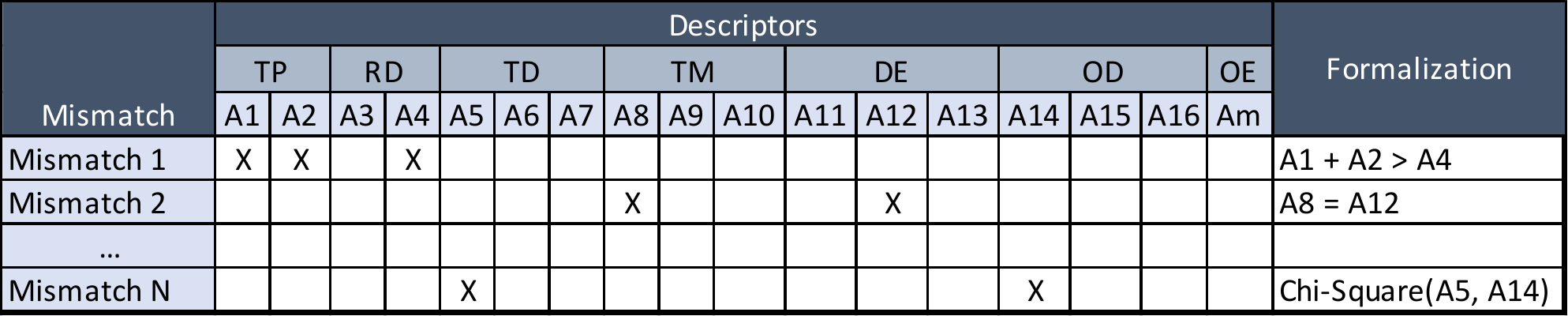}
  \caption{Mapping between Mismatches and ML-Enabled System Element Attributes} 
  \label{fig:mapping}
\end{figure}

As stated earlier, the end goal of our study is to use the ML mismatch information extracted from the interviews and survey responses to develop empirically-grounded machine-readable descriptors for different elements of ML-enabled systems. To this effect, in parallel we conducted a multi-vocal literature review \cite{Garousi2019} to identify software engineering best practices and challenges in the development and deployment of ML-enabled systems from both the academic and the practitioner perspective. Attributes for documenting elements of ML-enabled systems were extracted or inferred from the primary studies. In Phase 2 of our research, we will perform a mapping between system element attributes and mismatches, in which for each mismatch we identify the set of attributes that could be used to detect that mismatch, and formalize the mismatch as a predicate over those attributes, as shown in the Formalization column in Figure \ref{fig:mapping}. We will then perform gap analysis to identify mismatches that do not map to any attributes, and attributes that do not map to any mismatch. We then complement the mapping based on information from the interviews and survey plus our domain knowledge, by adding attributes and potentially adding new mismatches that could be detected based on the available attributes.

\begin{figure}[htbp]
  \centering
  \includegraphics[width=7cm]{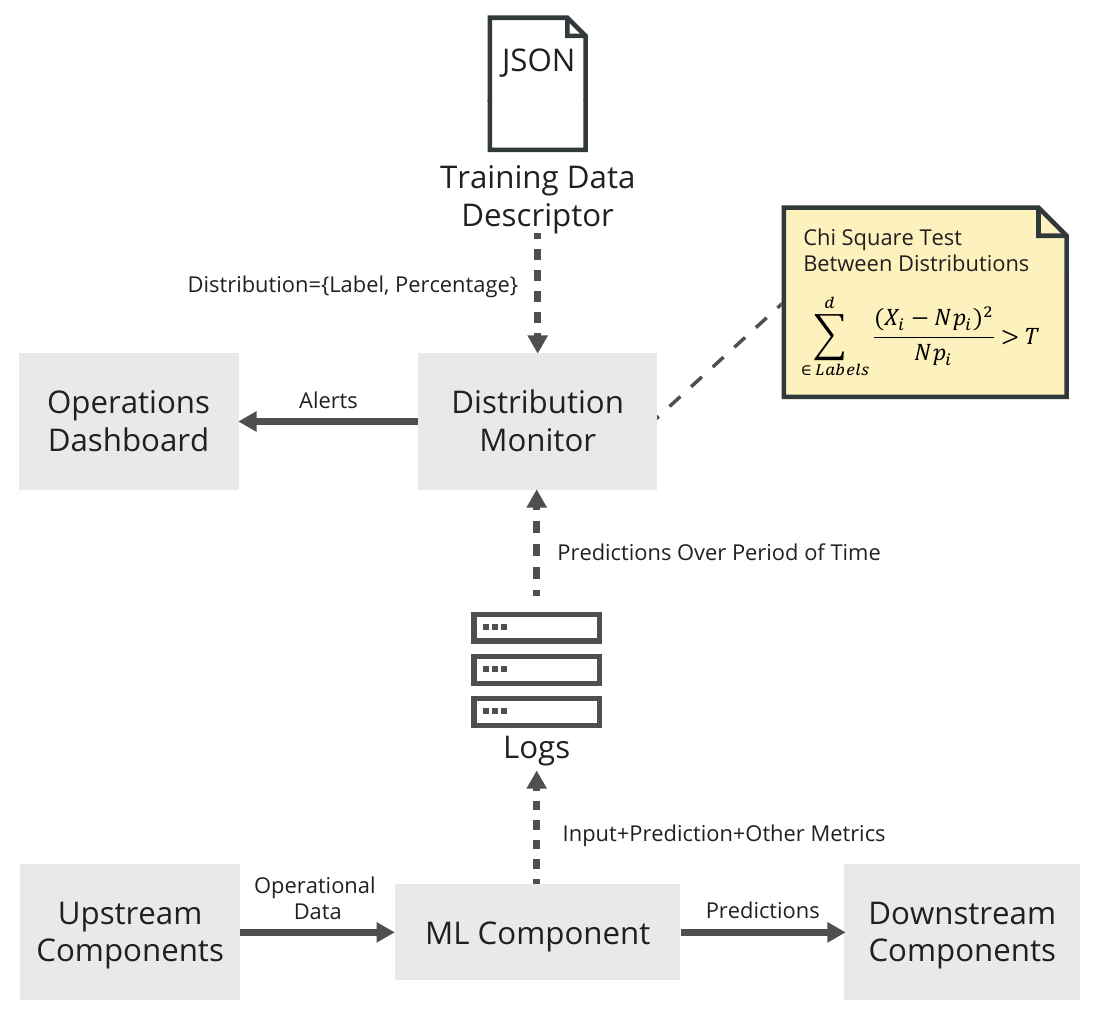}
  \caption{Data Drift Detection Tool} 
  \label{fig:tool}
\end{figure}

The resulting attributes will be codified into JSON Schema documents (\url{https://json-schema.org/}) that can be used by automated mismatch detection tools. These tools can range from a simple web-based client that reads in all descriptors and presents them to a user evaluating documentation, to a more elaborate tool or system component such as the one presented in Figure \ref{fig:tool}, which uses the \textit{Distribution} attribute from the \textit{Training Data} descriptor for runtime data drift detection, by performing a chi-square test between the distributions of training data and the operational data.

\section{Limitations}\label{sec:limitations}

Although the selected interviewees and survey participants have different affiliation, roles, and years of experience, as shown in Figures \ref{fig:demographics} and \ref{fig:survey-demographics}, we recognize that they might not be representative of all practitioners involved in the development and deployment of ML-enabled systems; therefore, the identified mismatches may not cover all potential mismatch types. We are also aware that the majority of interviewees and survey respondents are data scientists, and this might bias the results towards the data science perspective. To address this issue, in the interviews we always requested examples of mismatch caused by both information not provided and information not received, which gave us a broader set of mismatch examples, and indirectly included software engineer and operations as information providers. However, we fully recognize this limitation and propose to repeat the study at a larger scale and with a more balanced population. 
\section{Related Work}\label{sec:related-work}

Similar to our study, there are several practitioner-oriented studies that focus on software engineering challenges and best practices for the development and deployment of ML-enabled systems. Amershi et al \cite{Amershi2019} conducted a study with developers at Microsoft to understand how teams build software applications with customer-focused AI features. The results were a set of challenges and best practices, as well as the beginnings of a model of ML process maturity. Lwakatare et al \cite{Lwakatare2019} developed a taxonomy of software engineering challenges based on studying the development of ML systems in six companies. The challenges focused mostly on data science and were organized around a set of maturity stages related to the evolution of use of ML in commercial software-intensive systems. More recently, Serban et al \cite{Serban2020} conducted an academic and gray literature review of best practices for development of ML applications, and validated adoption in real projects via a practitioner survey. From the descriptor perspective, while there is existing, recent work in creating descriptors for data sets \cite{Bender2018}\cite{Gebru2018}\cite{Holland2018}, models \cite{Mitchell2019}, and online AI services \cite{Arnold2019}\cite{Richards2020}, there are three main limitations: (1) they do not address the software engineering and operations perspectives, (2) they are not machine-readable, and (3) they are targeted at selection or evaluation of existing data set and models and not at end-to-end system development. Our descriptors will address these three limitations.

\section{Conclusions and Next Steps}\label{sec:conclusions}
Empirically-validated practices and tools to support software engineering of ML-enabled systems are still in their infancy. In this paper, we presented the results of our study to understand the types of mismatch that occur in the development and deployment of ML-enabled systems due to incorrect assumptions made by different stakeholders. While best practices for ML model development are readily available, understanding how to deploy, operate, and sustain these models remains a challenge. The 7 categories of ML mismatch that we identified, along with their 34 subcategories, contribute to codifying the nature of the challenges. The Phase 1 results of our study demonstrate that improved communication and automation of ML mismatch awareness and detection can help improve software engineering of ML-enabled systems. The next steps of our study include developing the machine-readable descriptors as described in Section \ref{sec:towards-detection}, validating the descriptors in industry, and implementing their detection in tools such as the web-based descriptor viewer and the runtime data drift detector described in Section \ref{sec:towards-detection}. Our vision is to make the descriptors publicly available and create a community around tool development and descriptor extensions, with the end goal of improving the state of the engineering practices for development, deployment, operation, and evolution of ML-enabled systems.

\section*{Acknowledgments}
This material is based upon work funded and supported by the Department of Defense under Contract No. FA8702-15-D-0002 with Carnegie Mellon University for the operation of the Software Engineering Institute, a federally funded research and development center (DM20-0959).

\bibliographystyle{abbrv}
\bibliography{references}
\balance

\end{document}